\documentclass[preprint2,tighten]{aastex62}
\pdfoutput=1
\usepackage{amsmath,amstext}
\usepackage{apjfonts} 
\usepackage{url}

\newcommand{\unit}[1]{\ensuremath{\, \mathrm{#1}}}
\newcommand\kep{\emph{Kepler}}

\shorttitle{Kepler-503b: A Companion at the Hydrogen Burning Mass Limit}
\shortauthors{Ca\~nas et al. 2018}

\begin{document}
\title{Kepler-503b: An Object at the Hydrogen Burning Mass Limit Orbiting a Subgiant Star}

\correspondingauthor{Caleb I. Ca\~nas}
\email{canas@psu.edu}

\author{Caleb I. Ca\~nas}
\affiliation{Department of Astronomy \& Astrophysics, The Pennsylvania State University, 525 Davey Lab, University Park, PA 16802, USA}
\affiliation{Center for Exoplanets \& Habitable Worlds, University Park, PA 16802, USA}
\affiliation{Penn State Astrobiology Research Center, University Park, PA 16802, USA}

\author{Chad F. Bender}
\affiliation{Department of Astronomy and Steward Observatory, University of Arizona, Tucson, AZ 85721, USA}

\author{Suvrath Mahadevan}
\affiliation{Department of Astronomy \& Astrophysics, The Pennsylvania State University, 525 Davey Lab, University Park, PA 16802, USA}
\affiliation{Center for Exoplanets \& Habitable Worlds, University Park, PA 16802, USA}
\affiliation{Penn State Astrobiology Research Center, University Park, PA 16802, USA}

\author{Scott W. Fleming}
\affiliation{Space Telescope Science Institute, 3700 San Martin Dr., Baltimore, MD 21218, USA}

\author{Thomas G. Beatty}
\affiliation{Department of Astronomy \& Astrophysics, The Pennsylvania State University, 525 Davey Lab, University Park, PA 16802, USA}
\affiliation{Center for Exoplanets \& Habitable Worlds, University Park, PA 16802, USA}
\affiliation{Penn State Astrobiology Research Center, University Park, PA 16802, USA}

\author{Kevin R. Covey}
\affil{Department of Physics \& Astronomy, Western Washington University, Bellingham, WA 98225, USA}

\author{Nathan De Lee}
\affiliation{Department of Physics, Geology, and Engineering Technology, Northern Kentucky University, Highland Heights, KY 41099}
\affiliation{Department of Physics \& Astronomy, Vanderbilt University, Nashville, TN 37235}

\author{Fred R. Hearty}
\affiliation{Department of Astronomy \& Astrophysics, The Pennsylvania State University, 525 Davey Lab, University Park, PA 16802, USA}

\author{D. A. Garc\'ia-Hern\'andez}
\affiliation{Instituto de Astrof\'isica de Canarias (IAC), E-38205 La Laguna, Tenerife, Spain}
\affiliation{Universidad de La Laguna (ULL), Departamento de Astrof\'isica, E-38206 La Laguna, Tenerife, Spain}

\author{Steven R. Majewski}
\affiliation{Department of Astronomy, University of Virginia, Charlottesville, VA 22904, USA}

\author{Donald P. Schneider}
\affiliation{Department of Astronomy \& Astrophysics, The Pennsylvania State University, 525 Davey Lab, University Park, PA 16802, USA}
\affiliation{Center for Exoplanets \& Habitable Worlds, University Park, PA 16802, USA}

\author{Keivan G. Stassun}
\affiliation{Department of Physics \& Astronomy, Vanderbilt University, Nashville, TN 37235}

\author{Robert F. Wilson}
\affiliation{Department of Astronomy, University of Virginia, Charlottesville, VA 22904, USA}

\begin{abstract}
Using spectroscopic radial velocities with the APOGEE instrument and \emph{Gaia} distance estimates, we demonstrate that Kepler-503b, currently considered a validated {\it Kepler} planet, is in fact a brown-dwarf/low-mass star in a nearly circular 7.2-day orbit around a subgiant star. Using a mass estimate for the primary star derived from stellar models, we derive a companion mass and radius of \(0.075\pm0.003\unit{M_\odot}\) (\(78.6\pm3.1\) \unit{M_{Jup}}) and \(0.099^{+0.006}_{-0.004}\unit{R_\odot}\) (\(0.96^{+0.06}_{-0.04}\) \unit{R_{Jup}}), respectively. Assuming the system is coeval, the evolutionary state of the primary indicates the age is $\sim6.7$ Gyr. Kepler-503b sits right at the hydrogen burning mass limit, straddling the boundary between brown dwarfs and very low-mass stars. More precise radial velocities and secondary eclipse spectroscopy with James Webb Space Telescope will provide improved measurements of the physical parameters and age of this important system to better constrain and understand the physics of these objects and their spectra. This system emphasizes the value of radial velocity observations to distinguish a genuine planet from astrophysical false positives, and is the first result from the SDSS-IV monitoring of Kepler planet candidates with the multi-object APOGEE instrument.
\end{abstract}
\section{Introduction}
The NASA \kep{} space mission, in its search for transiting Earth analogues, provided nearly continuous observations of \(\sim 200,000\) stars with a photometric precision of a few parts per million \citep{Borucki2010,Koch2010}. The final \kep{} data release (DR25) lists more than \(8,000\) objects of interest (KOIs), or targets showing a transit which may be caused by an exoplanet \citep{Thompson2017}. Vetting these KOIs has revealed over \(2,000\) eclipsing binaries with precise photometric data \citep{Kirk2016}. While high-resolution imaging \citep{Furlan2017} and statistical methods \citep{Morton2016} are useful for constraining the nature of a KOI, dynamical observations can provide an unambiguous classification for a given system.

Eclipsing binaries are important astrophysical systems because simultaneous modeling of spectroscopic and photometric observations yields precise dynamical masses and stellar radii \citep{Torres2010}. Precisely measured stellar parameters are valuable for calibrating and refining stellar evolution models \citep[e.g.,][]{Fernandez2009,Torres2014} and even play a role in the cosmic distance scale. Furthermore, determining precise properties of exoplanets requires an understanding of their host stars' parameters, particularly the masses and radii. The detection of false positive KOIs has greater implications for the planet-hosting stellar population. The presence of eclipsing binaries quantifies the false positive rate and can reveal any dependencies on parameters, such as the location within the \kep{} field and properties of the host star or planetary candidate.

In this paper, we provide tight constraints on the age, radii, masses, and other properties of the low mass-ratio (\(M_{2}/M_{1}=q\sim 0.07\)) eclipsing binary system Kepler-503 (KIC 3642741, 2MASS J19223275+3842276, Kp = 14.75, H = 13.14). The paper is structured as follows: Section \ref{sec:observations} presents the observational data, Section \ref{sec:datared} discusses our data processing, and Section \ref{sec:models} describes our analysis. A discussion of our results is presented in Section \ref{sec:discussion}. 
\section{Observations}\label{sec:observations}
Kepler-503 was observed from Apache Point Observatory (APO) between 30 April 2015 and 7 April 2017 as part of the APO Galaxy Evolution Experiment (APOGEE) KOI program \citep{Fleming2015,Majewski2017,Zasowski2017} within SDSS-IV \citep{Blanton2017}. We obtained nineteen spectra of Kepler-503, using the high-resolution (\(R\sim22500\)), near-infrared (\(1.514-1.696\) \unit{\mu m}), multi-object APOGEE spectrograph \citep{Wilson2010,Wilson2012}, mounted on the Sloan 2.5-meter telescope \citep{Gunn2006}. The observations are summarized in Table \ref{tab:table1}, which lists our derived radial velocities, their uncertainties (corresponding to a \(1-\sigma\) error), and the signal-to-noise ratio (SNR) per pixel for each epoch. One observation with a SNR below \(10\) was not used for analysis.
\startlongtable
\begin{deluxetable*}{cccc}
\tablecaption{APOGEE Observations\label{tab:table1}}
\tablehead{\colhead{BJD\(_\text{TDB}^{\dagger}\)} & \colhead{RV (km s\(^{-1}\))} & \colhead{\(1-\sigma\) (km s\(^{-1}\))} & \colhead{SNR\(^\ddagger\)} (pixel\(^{-1}\))}
\startdata
2456932.708196 & -52.86 & 0.21 & 22 \\
2457142.860499 & -52.11 & 0.2 & 24 \\
2457199.743627 & -46.26 & 0.35 & 12 \\
2457264.658553 & -43.3 & 0.24 & 17 \\
2457265.795423 & -50.02 & 0.17 & 29 \\
2457266.783228 & -52.86 & 0.18 & 25 \\
2457294.705133 & -49.98 & 0.17 & 30 \\
2457319.648315 & -45.49 & 0.25 & 19 \\
2457472.004760 & -45.25 & 0.26 & 16 \\
2457498.977450 & -52.66 & 0.39 & 12 \\
2457527.949126 & -53.05 & 0.15 & 33 \\
2457554.814754 & -42.62 & 0.26 & 16 \\
2457562.871095 & -47.25 & 0.26 & 16 \\
2457643.740360 & -52.13 & 0.35 & 12 \\
2457672.675784 & -52.15 & 0.17 & 27 \\
2457701.553540 & -51.51 & 0.49 & 12 \\
2457831.992227 & -50.48 & 0.23 & 17 \\
2457850.990495 & * & * & 6\\
\enddata
\tablenotetext{\dagger}{BJD\(_\text{TDB}\) is the Barycentric Julian Date in the Barycentric Dynamical Time standard, which takes into consideration relativistic effects.}
\tablenotetext{\ddagger}{APOGEE has about two pixels per resolution element.}
\tablenotetext{*}{Value omitted for the observation with SNR < 10.}
\end{deluxetable*}
Kepler-503 was observed for the entirety of the \kep{} mission\added{. The transiting companion, Kepler-503b}, is listed as a planetary candidate in DR25\deleted{,} and was statistically validated as an exoplanet by \cite{Morton2016}. The final data release lists a shallow, \(0.37\%\) transit signal with an orbital period of \(7.258450123\) days and an estimated physical radius of \(5.53^{+1.59}_{-0.53}\) \unit{R_{\oplus}}. The stellar parameters in the DR25 stellar properties catalog \citep{Mathur2017} were derived with photometric priors and inferred from stellar models. The DR25 parameters for Kepler-503 suggest a solar-like host star with an effective temperature of \(5638^{+154}_{-171}\) K, a mass of \(1.006^{+0.090}_{-0.120}\)~\unit{M_{\odot}}, and a radius of \(0.920^{+0.264}_{-0.088}\)~\unit{R_{\odot}}.
\section{Data Processing}\label{sec:datared}
\subsection{Radial Velocities}\label{sec:reduc}
The APOGEE data reduction pipeline \citep{Nidever2015} performs sky subtraction, telluric and barycentric correction, and wavelength and flux calibration for each observation of a target. We focused our analysis on these individual spectra for dynamical characterization. While the APOGEE pipeline provides radial velocity measurements, we performed additional post-processing on the spectrum to remove residual telluric lines prior to analysis. 

Radial velocities of Kepler-503 were derived using the cross-correlation method with uncertainties calculated via the maximum-likelihood approach presented by \cite{Zucker2003}. With this method, we account for uncertainty contributions from the spectral bandwidth, sharpness of the correlation peak, and the spectral line SNR but cannot exclude systematic uncertainties due to the instrument or poor template selection. Cross-correlation searches for the Doppler shift of an observed spectrum that maximizes the correlation with a template that adequately represents the observed spectrum without a Doppler shift. The best-fitting spectrum should have the highest correlation and it is common practice for this to be used as the template \cite[e.g.,][]{Latham2002}. We identified the best-fitting synthetic spectrum in the H-band from a grid of BT-Settl synthetic spectra \citep{Allard2012} by cross-correlating the APOGEE epoch with the highest SNR against a grid spanning surface effective temperature (\(5100\le T_{e}\le6100\), in intervals of 100 K), surface gravity (\(3.5\le\log g\le4.5\), in intervals of 0.5 dex), metallicity (\(-0.5\le\text{[M/H]}\le0.5\), in intervals of 0.5 dex), and rotational broadening (\(2 \le v\sin i\le 50\), in intervals of 2 \unit{km s^{-1}}). The synthetic spectrum with the largest correlation was used for the final cross-correlation to derive the reported radial velocities in Table \ref{tab:table1}. The properties of the best model are listed in Table \ref{tab:table2} and were not used as priors for fitting the system.
\subsection{Photometry}
We used the \kep{} pre-search data conditioned time-series light curves \citep{Smith2012,Stumpe2012,Stumpe2014} available at the Mikulski Archive for Space Telescopes (MAST). We assumed the transit signal was superimposed on the stellar variability and could be modeled using a Gaussian process. We used the {\tt celerite} package and assumed a quasi-periodic covariance function for the Gaussian process, following the procedure in \cite{Foreman-Mackey2017}. No additional processing was performed on the light curve.
\section{Results}\label{sec:models}
\begin{figure*}[!ht]
\epsscale{1.15}
\plotone{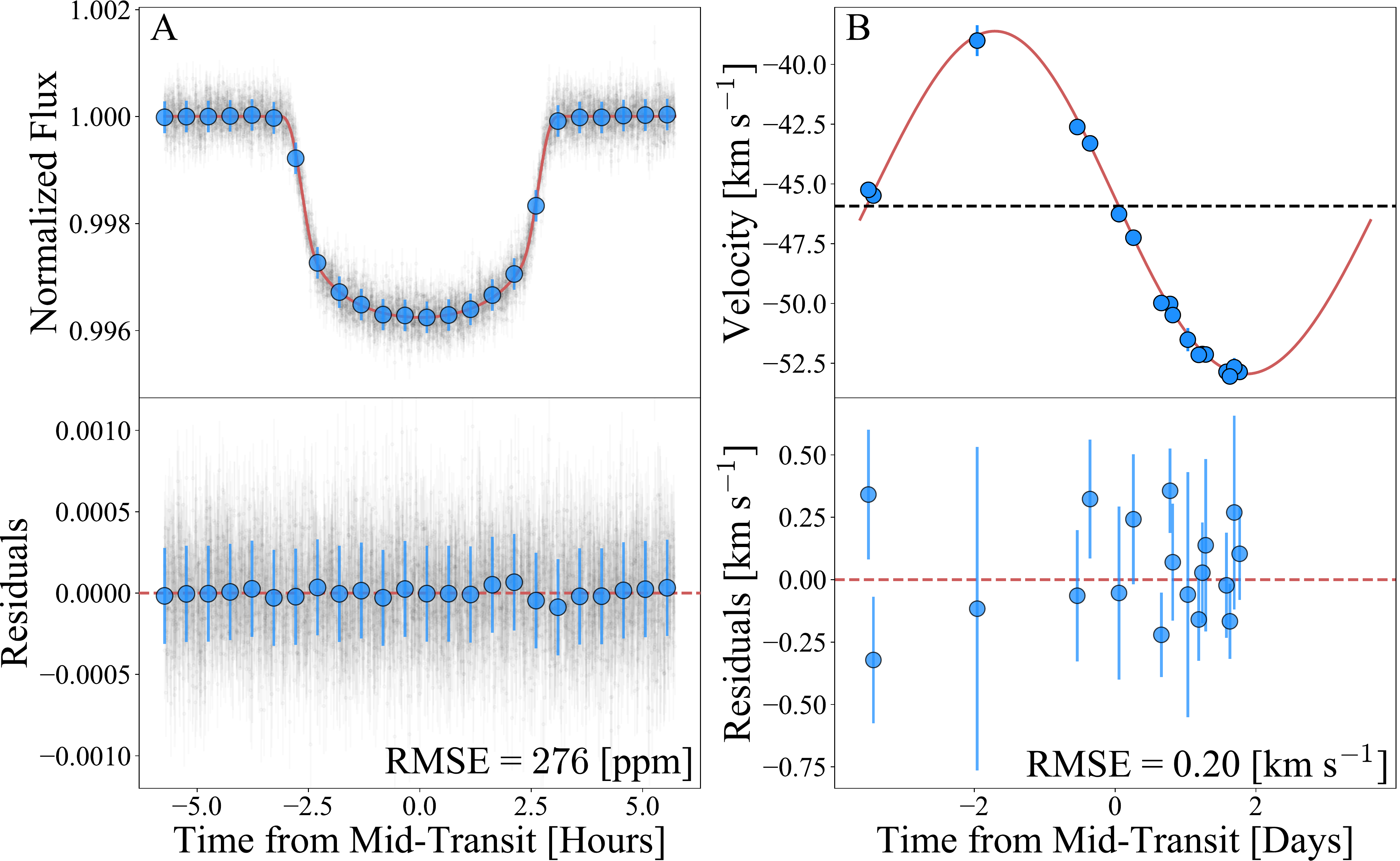}
\caption{The photometry and velocimetry of Kepler-503. Panel A displays the phase-folded photometry from DR25. The small dots are the raw data while the larger circles are binned to the \kep{} long-cadence. Panel B shows the radial velocities from Table \ref{tab:table1} phased to the period of the system. In each case, the top panel contains the model as a solid line while the bottom panel shows the residuals and the root mean square error. The derived mass and radius ratios are \(M_{2}/M_{1}=0.0648\) and \(R_{2}/R_{1}=0.05619\), respectively. The modeled stellar parameters for the host star, Kepler-503A, give properties of the secondary companion that are incompatible with that of an exoplanet. \label{fig:f1}}
\end{figure*}

We jointly modeled Kepler-503's radial velocities and light curve using the {\tt EXOFASTv2} analysis package \citep{Eastman2017}. The light curve and Keplerian radial velocity models follow the parametrization described by \cite{Eastman2013}. We adopted a quadratic limb darkening law for the transit. The priors for the modeling included (i) 2MASS \(JHK\) magnitudes \citep{Skrutskie2006}, (ii) SDSS \(ugriz\) magnitudes \citep{Alam2015}, (iii) \(UBV\) magnitudes \citep{Everett2012}, (iv) WISE magnitudes \citep{Wright2010}, (v) surface gravity, temperature and metallicity from the APOGEE Stellar Parameter and Chemical Abundances Pipeline \citep[ASPCAP,][]{GarciaPerez2016}, (vi) the maximum visual extinction from estimates of Galactic dust extinction by \cite{Schlafly2011}, (vii) the photometric measurements from the second data release of the \emph{Gaia} survey \citep{GaiaCollaboration2018}, and (viii) the distance estimate from \cite{Bailer-Jones2018}. We validated the performance of our {\tt EXOFASTv2} implementation by analyzing the corresponding data products for KOI-189, and obtained very similar parameters to those published by \cite{Diaz2014} using the {\tt PASTIS} planet-validation software. We repeated the analysis for Kepler-503 using the parallax from \emph{Gaia} to determine if there were any significant effects from the nonlinearity of the parallax transformation or any asymmetry in the parallax posterior distribution. The results are consistent to within their \(1-\sigma\) uncertainties and herein we present values from the analysis using the distance prior. Figure \ref{fig:f1} presents the result of the fit and Table \ref{tab:table2} provides a summary of the stellar priors and the inferred systemic parameters along with their confidence intervals. The minimum companion mass is \(\sim 0.075\) \unit{M_{\odot}}, a value incompatible with the statistical validation and estimated parameters from DR25.
\startlongtable
\begin{deluxetable*}{llc}
\tablecaption{Parameters for the Kepler-503 System \label{tab:table2}}
\tablehead{\colhead{~~~Parameter} & \colhead{Units} & \colhead{Median Value}}
\startdata
\sidehead{Primary Synthetic Spectrum\(^\dagger\):}
~~~Effective Temperature \dotfill & $T_{e}$ (K)\dotfill & $6000$\\
~~~Surface Gravity \dotfill & $\log(g_1)$ (cgs)\dotfill & $4.0$\\
~~~Metallicity \dotfill & [M/H]\dotfill & $0.0$\\
~~~Rotational Velocity \dotfill & \(v\sin i\) (km \unit{s^{-1}})\dotfill & $2.0$\\
\sidehead{Primary Stellar Priors:}
~~~Effective Temperature\(^\ddagger\) \dotfill & $T_{e}$ (K)\dotfill & $5690 \pm 150$\\
~~~Surface Gravity\(^\ddagger\) \dotfill & $\log(g_1)$ (cgs)\dotfill & $4.0$\\
~~~Metallicity\(^\ddagger\) \dotfill & [Fe/H]\dotfill & $0.17\pm 0.05$\\
~~~Maximum Visual Extinction \dotfill & \(A_{V,max}\) \dotfill & $0.43$\\
~~~Distance\dotfill & (pc)\dotfill & $1628 \pm 55$\\
\sidehead{Primary Parameters:}
~~~Mass \dotfill & $M_{1}$ (\unit{M_{\odot}})\dotfill & $1.154^{+0.047}_{-0.042}$\\
~~~Radius\dotfill & $R_{1}$ (\unit{R_{\odot}})\dotfill & $1.764^{+0.080}_{-0.068}$\\
~~~Density \dotfill & $\rho_1$ (g \unit{cm^{-3}})\dotfill & $0.297^{+0.038}_{-0.037}$\\
~~~Surface Gravity \dotfill & $\log(g_1)$ (cgs)\dotfill & $4.008 \pm 0.038$\\
~~~Effective Temperature\dotfill & $T_{e}$ (K)\dotfill & $5670^{+100}_{-110}$\\
~~~Metallicity\dotfill & [Fe/H]\dotfill & $0.169^{+0.046}_{-0.045}$\\
~~~Age\dotfill & (Gyr)\dotfill & $6.7^{+1.0}_{-0.9}$\\
~~~Parallax\dotfill & (mas)\dotfill & $0.617^{+0.020}_{-0.019}$\\
\sidehead{Secondary Parameters:}
~~~Mass\dotfill & $M_{2}$ (\unit{M_{\odot}})\dotfill & $0.075\pm0.003$\\
~~~Radius\dotfill & $R_{2}$ (\unit{R_{\odot}})\dotfill & $0.099_{-0.004}^{+0.006}$\\
~~~Density\dotfill & $\rho_{2}$ (g \unit{cm^{-3}})\dotfill & $108\pm17$\\
~~~Surface Gravity\dotfill & $\log(g_{2})$ (cgs)\dotfill & $5.320^{+0.045}_{-0.050}$\\ 
~~~Equilibrium Temperature\dotfill & $T_{eq}$ (K)\dotfill & $1296^{+22}_{-23}$\\
\sidehead{Orbital Parameters:}
~~~Orbital Period\dotfill & $P$ (days) \dotfill& $7.2584481 \pm 0.0000023$\\
~~~Time of Periastron\dotfill & $T_{P}$ (BJD\(_{\text{TDB}}\))\dotfill & $2454970.27^{+0.63}_{-1.1}$\\
~~~Semi-major Axis\dotfill & $a$ (AU) \dotfill& $0.0786\pm0.0010$\\
~~~Orbital Eccentricity\dotfill & $e$ \dotfill& $0.025^{+0.014}_{-0.012}$\\
~~~Argument of Periastron\dotfill & $\omega$ (degrees)\dotfill & $33^{+32}_{-54}$\\
~~~Semi-amplitude Velocity\dotfill & $K$ (km \unit{s^{-1}})\dotfill & $7.17\pm0.18$\\
~~~Mass Ratio\dotfill & $q$ \dotfill & $0.0648^{+0.0019}_{-0.0020}$\\
~~~Systemic Velocity\dotfill & $\gamma$ (km \unit{s^{-1}})\dotfill & $-45.93\pm0.10$\\
~~~Radial Velocity Jitter\dotfill & $\sigma_{RV}$ (m \unit{s^{-1}})\dotfill & $168^{+96}_{-92}$\\
\sidehead{Transit Parameters:}
~~~Time of Mid-transit\dotfill & $T_C$ (BJD\textsubscript{TDB})\dotfill & $2454971.34494 \pm 0.00026$\\
~~~Radius Ratio\dotfill & $R_{2}/R_{1}$ \dotfill & \(0.05619^{+0.00069}_{-0.00060}\)\\
~~~Scaled Semi-major Axis\dotfill & $a/R_{1}$ \dotfill & $9.59^{+0.39}_{-0.42}$\\
~~~e\(\cos\omega\)\dotfill & \dotfill & $0.017^{+0.009}_{-0.010}$\\
~~~e\(\sin\omega\)\dotfill & \dotfill & $0.010^{+0.020}_{-0.015}$\\
~~~Linear Limb-darkening Coefficient\dotfill & $u_1$\dotfill & $0.419 \pm 0.023$\\
~~~Quadratic Limb-darkening Coefficient\dotfill & $u_2$\dotfill & $0.238^{+0.044}_{-0.045}$\\
~~~Orbital Inclination\dotfill & $i$ (degrees)\dotfill & $88.04^{+1.0}_{-0.76}$\\
~~~Impact Parameter\dotfill & $b$\dotfill & $0.32^{+0.11}_{-0.17}$\\
\enddata
\tablenotetext{\dagger}{These parameters are for the best-fit template spectrum used for cross-correlation.}
\tablenotetext{\ddagger}{These values are from ASPCAP.}
\end{deluxetable*}
The stellar parameters from {\tt EXOFASTv2} are derived using stellar models; the values suggest Kepler-503 is an evolved star beyond the terminal age main sequence. \cite{Seager2003} proposed a diagnostic for a transiting system using transit parameters to obtain an estimate of the primary stellar density (\(\rho_{1}\)). With both photometry and velocimetry, one can determine that the observational data are consistent with the selected stellar models \citep[e.g.,][]{vonBoetticher2017}. To determine if the deviation from the DR25 stellar parameters was justified, we used the MESA Isochrones \& Stellar Tracks \citep[MIST,][]{Dotter2016,Choi2016} to model the primary star and iteratively refine its parameters until the density derived from the photometry and velocimetry was within the uncertainty of the density from the models. 

\begin{figure*}[!ht]
\epsscale{1.00}
\plottwo{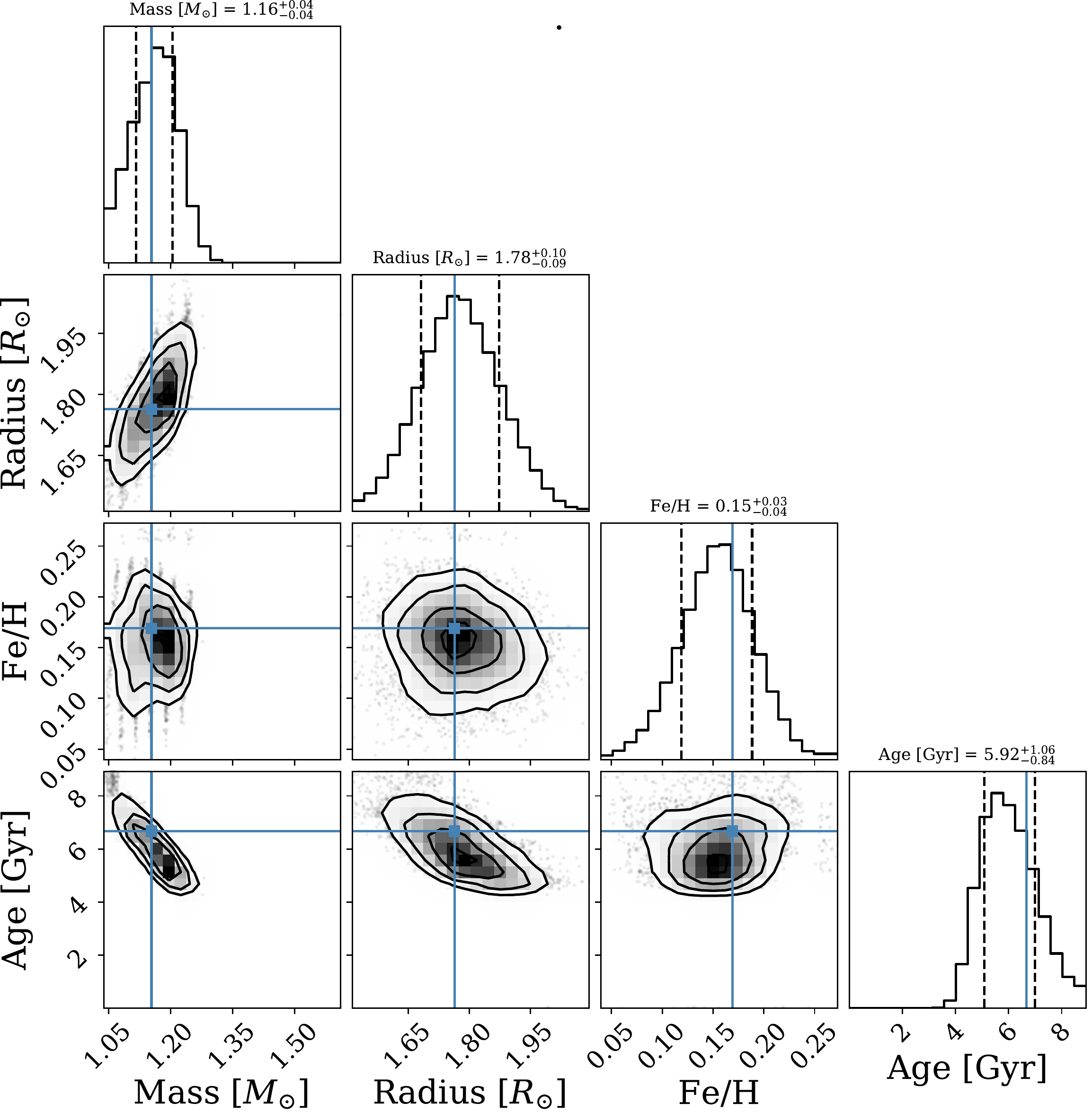}{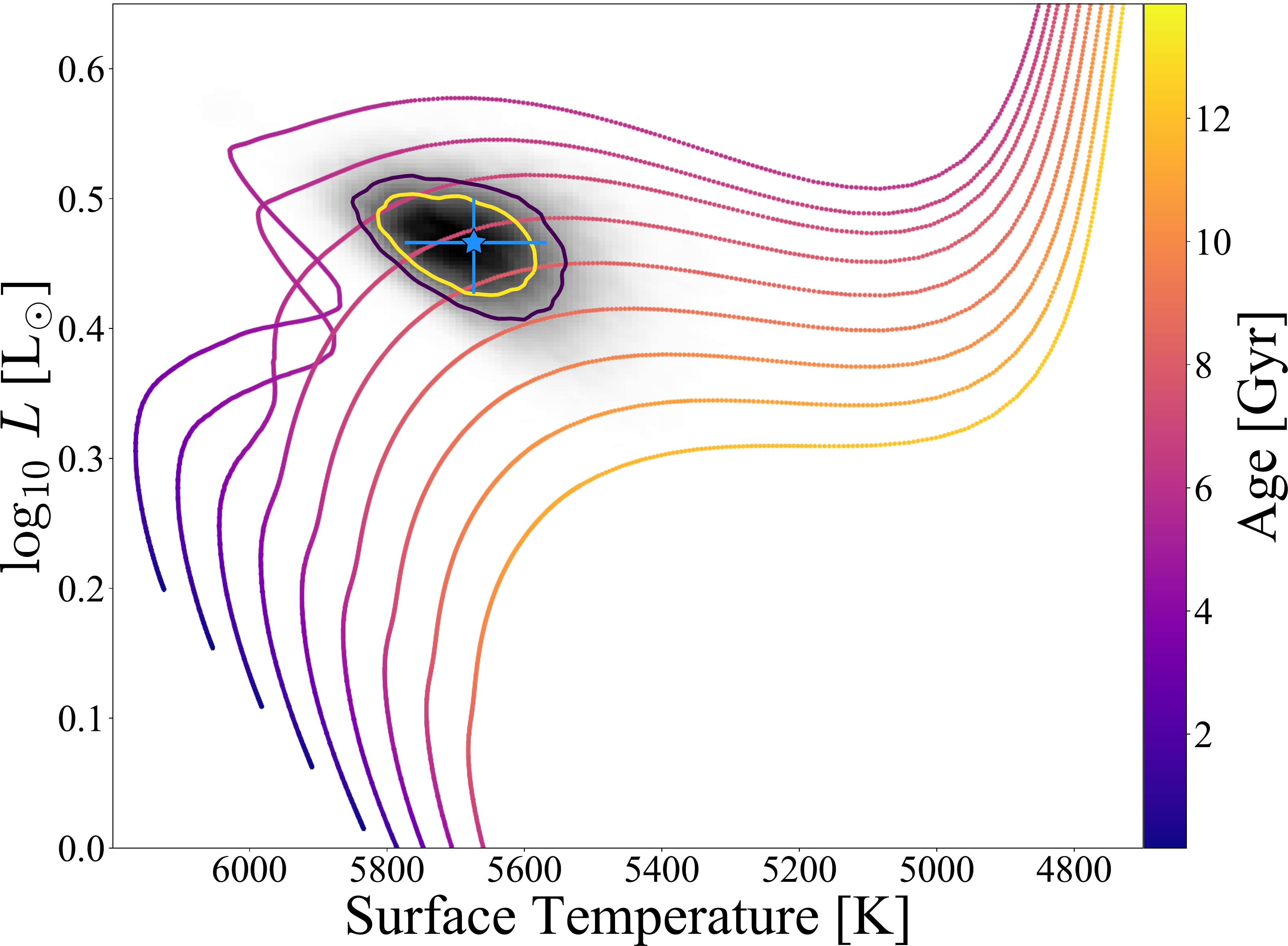}
\caption{The stellar parameters for the host star Kepler-503A. The left panel shows the posterior distribution for physical parameters of Kepler-503A after modeling various photometric parameters with MIST models. The values derived from {\tt EXOFASTv2} using the Yale-Yonsei isochrones are marked with lines for clarity. The models suggest Kepler-503A is an evolved subgiant star. The right panel shows the placement of the primary star on a Hertzsprung-Russel Diagram. The two-dimensional kernel density estimate of the posterior distribution from the {\tt EXOFASTv2} fit is included for reference. MIST evolution tracks are indicated for masses between 1.0 and 1.2 \unit{M_{\odot}} in steps of 0.025 \unit{M_{\odot}} and are shaded according to age. The placement of Kepler-503A is consistent with a star near the subgiant branch. \label{fig:f2}}
\end{figure*}

The result of the stellar modeling is shown in Figure \ref{fig:f2} and demonstrates that the stellar parameters are comparable to those derived with {\tt EXOFASTv2}. The data demonstrate that the primary star, Kepler-503, is not a solar analogue, but instead a slightly evolved subgiant. When compared to stellar evolution models, the posterior distributions for the primary star surface effective temperature and luminosity are located in the subgiant branch, which is in agreement with a slightly evolved system. Accordingly, these stellar parameters suggest that the purported exoplanet is actually an object near the hydrogen burning \added{mass} limit of \(\sim0.075\) \unit{M_{\odot}} \citep[e.g.,][]{Chabrier2005}.  
\section{Discussion}\label{sec:discussion}
\subsection{Constraint on the Companion Age}
Kepler-503, and particularly the secondary component, is of great astrophysical interest because the age of a star is a fundamental parameter that is often poorly constrained. This situation forces age estimates to rely on proxies, such as magnetic activity, element depletion, rotation \citep[gyrochronology, e.g.,][]{Soderblom2010}, or asteroseismology \citep[e.g.,][]{Pinsonneault2018}, which are difficult to measure for \added{low-mass} objects like Kepler-503b. This measurement is further complicated in low-mass stars because their spindown timescales can exceed the age of the Galaxy \citep{West2008}, and their fully convective nature can deplete lithium after only a few hundred million years \citep{Stauffer1998}. Objects near the bottom of the stellar mass function are of interest because they define the transition from bona fide stars to planets. The simplest way to determine the age of such an object is to associate it with another star or group for which the age is better constrained. Here, we assume Kepler-503 is a coeval system. The subgiant nature of the primary star places a strong constraint on the age of the companion because the placement of the subgiant branch on the Hertzsprung-Russell diagram is allowed for only a limited age range, according to stellar evolution models \citep[e.g.,][]{Soderblom2010}. The modeling suggests the age of the Kepler-503 system is \(6.7^{-0.9}_{+1.0}\) Gyr, making it considerably older than the Solar System. 
\subsection{Constraints on the Companion Temperature}
The lack of a secondary eclipse in the photometry provides additional information about Kepler-503b. The dearth of an eclipse, despite the low eccentricity and inclination of the orbit, places a constraint on the companion's effective surface temperature. The estimated eclipse depth is a function of both the radius ratio, \(R_{2}/R_{1}\), and the surface brightness ratio, \(B_{2}/B_{1}\). Using the parameters derived in this paper, the equilibrium temperature of Kepler-503b is \(\sim1296\) K. For such an object, the estimated transit depths for \emph{Spitzer}'s 3.6 and 4.5 \unit{\mu m} band-passes are \(\sim150\) and \(\sim220\) ppm, respectively. Even if we were to assume a higher surface effective temperature that is appropriate for a very low-mass star (\(\sim 2300\) K), the estimated eclipse depths for Kepler-503 are at the sensitivity limits of these \emph{Spitzer} channels. For \kep{}, the estimated eclipse depth at 600 nm is \(< 1\) ppm and is dwarfed by the variance in the light curve. Future photometric and spectroscopic observations with the James Webb Space Telescope are required to further constrain the flux ratio of the system to better characterize Kepler-503b. 
\subsection{Constraints on Evolutionary Models}
\begin{figure*}[!ht]
\epsscale{1.15}
\plotone{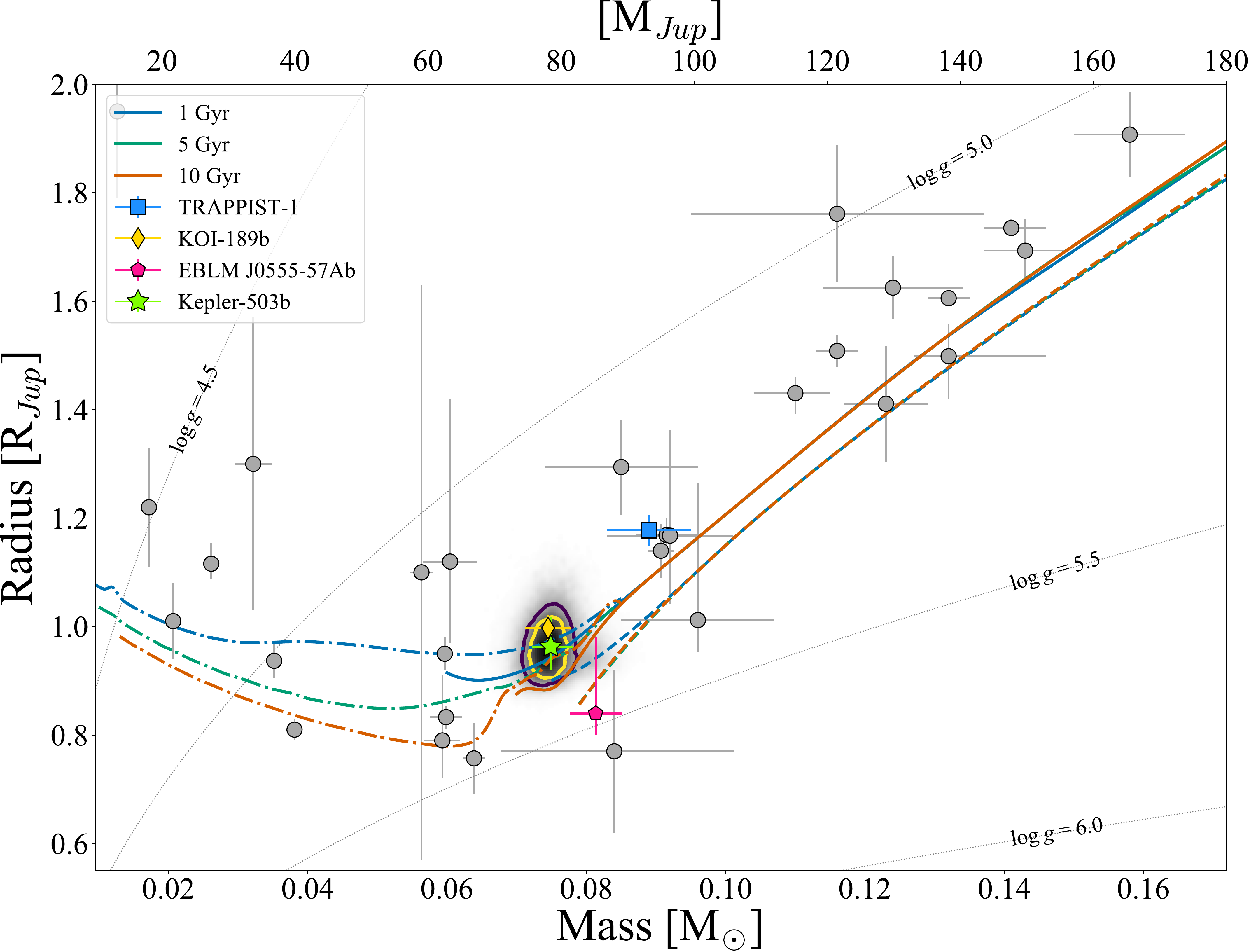}
\caption{The mass and radius posterior distribution of the companion, Kepler-503b. Kepler-503b is not an Earth-size planet but an object near the hydrogen burning mass limit. The parameters derived for Kepler-503b are plotted along a two-dimensional kernel density estimate of the posterior distribution from our fit. The contours represent the \(1-\sigma\) and \(2-\sigma\) (\(\sim 39.3\%\) and \(\sim 63.2\%\) of the volume, respectively). For reference, the very low-mass star TRAPPIST-1\replaced{and}{,} the brown dwarf KOI-189b\added{, and the low-mass star EBLM J0555-57Ab} are plotted as a square\replaced{ and diamond}{, diamond, and pentagon}, respectively, while other low-mass stars and high-mass sub-stellar companions from the literature appear as circles \cite[see][]{vonBoetticher2017}. Lines of constant surface gravity are drawn and masses are given in both solar and Jovian units. Models for sub-stellar companions and low-mass stars span the ages \(1,5,\text{ and }10\) Gyr. The solid lines are evolutionary tracks from \cite{Baraffe2015} assuming solar metallicity (\([\text{M/H}]=0.0\)), the dashed lines are from \cite{Baraffe1998} assuming sub-solar metallicity (\([\text{M/H}]=-0.5\)), and the dash-dotted lines are from \cite{Saumon2008} for L- and T-dwarfs assuming solar metallicity and a hybrid cloudy/cloudless evolutionary sequence. This figure is adapted from \cite{vonBoetticher2017}. \label{fig:f3}}
\end{figure*}

The study of objects similar to Kepler-503b is critical for the empirical calibration of the mass-radius relation near the hydrogen burning mass limit (\(\sim 0.075\) \unit{M_{\odot}}). Figure \ref{fig:f3} shows posterior distribution of Kepler-503b on a mass-radius diagram for brown dwarfs and low-mass stars. Its mass and radius are comparable to those of the brown dwarf KOI-189b \citep{Diaz2014} \replaced{, with values that span the boundary between L- and M- dwarfs. }{and the metal-poor ([Fe/H]=\(-0.24\pm0.16\)), low-mass star EBLM J0555-57Ab \citep{vonBoetticher2017}, with values that span the boundary between L- and M- dwarfs. The mass of both KOI-189b and EBLM J0555-57Ab are within one standard deviation from Kepler-503b, but the estimated ages are very different. KOI-189b has an estimated age comparable to the age derived for the Kepler-503 system of \(6.9^{+6.4}_{-3.4}\) Gyrs while EBLM J0555-57Ab is estimated to be much younger at \(1.9\pm1.2\) Gyr. The evolutionary tracks in Figure \ref{fig:f3}, show the change in metallicity cannot account for the change in radius of EBLM J0555-57Ab. We expect a population of objects spanning the hydrogen burning mass limit at varying masses and ages, but this population is still poorly characterized, primarily due to the extreme paucity of well-measured objects.}

Kepler-503b is one of the few objects in the regime where evolutionary tracks converge \added{with a well-constrained age}, and can thus help refine said models. The properties of Kepler-503b appear consistent with evolutionary tracks for solar metallicity low-mass stars and brown dwarfs. From the L- and T-dwarf models by \cite{Saumon2008}, the 5 and 10 Gyr evolutionary tracks are the best matches for this object. One caveat is that, while the derived metallicity for the host star suggests this is a slightly metal-rich system, the cloud-based models exist only for solar metallicities. Given the recent interest in very low-mass stars as targets for exoplanet searches \citep[e.g.,][]{Gillon2017}, it is essential that the super-solar metallicity regime be properly characterized. Future searches for planets around ultra-cool dwarfs will require precise masses and radii of the host star to properly characterize any detected exoplanets.
\section{Summary}\label{sec:summ}
This paper reveals a low mass-ratio eclipsing binary system in a nearly circular orbit that was erroneously classified as a transiting exoplanet. Analysis of the photometric and spectroscopic data shows Kepler-503 is an old system with a companion at the hydrogen burning mass limit. This misclassification is largely due to the stellar parameters previously adopted (i.e., a solar-like host). The stellar classification from DR25 used a prior which is known to underestimate the number of sub-giants due to Malmquist bias \citep[e.g.,][]{Bastien2014}. \cite{Mathur2017} acknowledge that some systematic biases persist in the DR25 stellar properties catalog, resulting in misclassified systems such as Kepler-503.

This study is one example of the systems observed in the ongoing APOGEE KOI program, with the ultimate goal of (i) refining the false positive rate of \kep{} exoplanet candidates, (ii) revealing any dependencies with stellar or candidate parameters, and (iii) understanding binarity and its effect in the planet host population. The recent second data release from the \emph{Gaia} survey will be helpful for future validation of KOIs by providing a parallax that help to constrain the properties of the host star. 
\acknowledgments
\added{We thank the anonymous referee for comments that improved the quality of this publication.}
CIC, CFB, and SM acknowledge support from NSF award AST 1517592.
ND, SRM, and RFW would like to acknowledge support from NSF Grant Nos. 1616684 and 1616636.
DAGH acknowledges support provided by the Spanish Ministry of Economy and Competitiveness (MINECO) under grant AYA-2017-88254-P.
Some of the data presented in this paper were obtained from MAST. STScI is operated by the Association of Universities for Research in Astronomy, Inc., under NASA contract NAS5-26555. Support for MAST for non-HST data is provided by the NASA Office of Space Science via grant NNX09AF08G and by other grants and contracts. 2MASS is a joint project of the University of Massachusetts and IPAC at Caltech, funded by NASA and the NSF.
Funding for the \kep{} mission is provided by the NASA Science Mission directorate. The NASA Exoplanet Archive is operated by Caltech, under contract with NASA under the Exoplanet Exploration Program.

Funding for SDSS-IV has been provided by the Alfred P. Sloan Foundation, the U.S. Department of Energy Office of Science, and the Participating Institutions. SDSS-IV acknowledges support and resources from the Center for High-Performance Computing at the University of Utah. The SDSS web site is \url{www.sdss.org}. SDSS-IV is managed by the Astrophysical Research Consortium for the Participating Institutions of the SDSS Collaboration including the Brazilian Participation Group, the Carnegie Institution for Science, Carnegie Mellon University, the Chilean Participation Group, the French Participation Group, Harvard-Smithsonian Center for Astrophysics, Instituto de Astrof\'isica de Canarias, The Johns Hopkins University, Kavli Institute for the Physics and Mathematics of the Universe (IPMU) / University of Tokyo, Lawrence Berkeley National Laboratory, Leibniz Institut f\"ur Astrophysik Potsdam (AIP), Max-Planck-Institut f\"ur Astronomie (MPIA Heidelberg), Max-Planck-Institut f\"ur Astrophysik (MPA Garching), Max-Planck-Institut f\"ur Extraterrestrische Physik (MPE), National Astronomical Observatories of China, New Mexico State University, New York University, University of Notre Dame, Observat\'ario Nacional / MCTI, The Ohio State University, Pennsylvania State University, Shanghai Astronomical Observatory, United Kingdom Participation Group, Universidad Nacional Aut\'onoma de M\'exico, University of Arizona, University of Colorado Boulder, University of Oxford, University of Portsmouth, University of Utah, University of Virginia, University of Washington, University of Wisconsin, Vanderbilt University, and Yale University.
%

%
\listofchanges
\end{document}